\documentstyle[12pt]{article}
\newcommand{\alphas}{\alpha_{\scriptscriptstyle S}}
\newcommand{\jpsi}{J/\psi}

\newcommand{\mochij}{\left\langle{\cal O}^{\chi_J}_8({}^3S_1)\right\rangle}

\newcommand{\mochio}{\left\langle{\cal O}^{\chi_1}_8({}^3S_1)\right\rangle}

\newcommand{\mojpsa}{\left\langle{\cal O}^{J/\psi}_8({}^3S_1)\right\rangle}
\newcommand{\mojpsb}{\left[
          \left\langle{\cal O}^{J/\psi}_8({}^1S_0)\right\rangle
    +{\displaystyle3\over\displaystyle m^2}
          \left\langle{\cal O}^{J/\psi}_8({}^3P_0)\right\rangle
    +{\displaystyle4\over\displaystyle5m^2}
          \left\langle{\cal O}^{J/\psi}_8({}^3P_2)\right\rangle\right]}

\begin{document}
\renewcommand{\thefootnote}{\fnsymbol{footnote}}
\begin{flushright}
TIFR/TH/96-29\\
\end{flushright}
\vspace*{3cm}
\begin{center}
{\Large \bf \boldmath Charmonium Production \footnote{Presented at the 
XXXI Rencontres de Moriond, ``QCD and High Energy Hadronic 
Interactions'', March 23-30,  1996, Les Arcs, France.}} \\
\vskip 32pt
{\bf K. Sridhar\footnote{sridhar@theory.tifr.res.in}}\\
{\it Theory Group, Tata Institute of Fundamental Research, \\ 
Homi Bhabha Road, Bombay 400 005, India.}

\vspace{125pt}
{\bf ABSTRACT}
\end{center}
\vspace{12pt}

Theoretical analyses of the results on quarkonium production 
at large transverse momentum ($p_T$) in $p\bar p$ collisions, 
at the Tevatron have revealed two novel features of
the physics of quarkonium production : 1) the contribution of
fragmentation of gluons and charm quarks to the cross-section 
at large $p_T$, and 2) the importance of colour octet
components of the quarkonium wave-function. I discuss the
theoretical developments which have contributed to a
reasonably consistent picture of quarkonium production at
the Tevatron, and discuss large-$p_T$ $J/\psi$ production
at HERA and at LHC as important tests of the fragmentation
picture. I also discuss our recent analysis of 
$J/\psi$ production cross-sections at fixed-target 
energies, where we find that the energy dependence of the
$p_T$-integrated cross-sections for both $pp$ and $\pi^- - p$ collisions
is reasonably well reproduced, when the colour-octet components
are included.
\vspace{98pt}
\noindent
\begin{flushleft}
June 1996\\
\end{flushleft}

\setcounter{footnote}{0}
\renewcommand{\thefootnote}{\arabic{footnote}}

\vfill
\clearpage
\pagestyle{empty}
\baselineskip 16pt
Quarkonium production has conventionally been described in
the colour-singlet model \cite{berjon, br}, wherein a
heavy-quark pair produced via parton-fusion processes is projected 
onto a physical quarkonium state using a colour-singlet projection and 
an appropriate spin-projection. This model has been successfully applied 
\cite{br} to describe large-$p_T$ $J/\psi$ production in the ISR experiment. 
However, the inclusive $J/\psi$ production cross-section measured by the
CDF experiment at the Tevatron \cite{cdf} turned out to be an order 
of magnitude larger than the prediction of the colour-singlet model. 

It was realised \cite{bryu} that in addition to the 
parton fusion contributions, fragmentation of gluons and charm quarks 
could be an important source of large-$p_T$ $J/\psi$ production at
high energies. This is 
computed by factorising the cross-section for the process 
$AB \rightarrow (J/\psi,\chi_i) X$ (where $A,\ B$ denote hadrons)
into a part containing the hard-scattering cross-section for producing a
parton of large-$p_T$ but zero virtuality,
and a part which specifies the fragmentation of
the gluon or the charm quark into the required charmonium state. 
The fragmentation function can be computed
perturbatively, in the same spirit as in the colour-singlet model.
This yields the fragmentation function at an initial scale $\mu_0$
which is of the order of $m_c$, and large logarithms in $p_T/m_c$ 
which appear are resummed using the Altarelli-Parisi equation.
The gluon and charm fragmentation functions 
have been calculated \cite{bryu, frag}
and using these inputs it has been found \cite{jpsi} that
the order-of-magnitude discrepancy between the 
theory and the CDF data can be resolved. 

Another important aspect of the physics of quarkonia revealed by 
the analyses of the CDF data is the importance of colour-octet 
contributions. A systematic formulation based on non-relativistic QCD, 
using the factorisation method has been recently carried out \cite{bbl}, and 
in this formulation the quarkonium wave-function admits of a 
Fock-space expansion in powers of $v$, the relative velocity between
the heavy quarks; for example, the $\chi$ states have the 
colour-singlet $P$-state component at leading order, but there exist 
additional contributions at non-leading order in $v$, which involve octet 
$S$-state components. The octet component allows a consistent
perturbative treatment of $\chi$ decays \cite{bbl2}, whereby the
infrared divergence appearing in the colour-singlet decay amplitudes 
\cite{bgr} can be absorbed via a wave-function renormalisation. 
As in the case of decays, the $P$-state fragmentation functions are 
also infra-red divergent and, hence, they include the 
octet component.

For $S$-state resonances, the octet contribution is suppressed by powers of
$v$. Further, the $S$-wave amplitude is not infrared divergent. 
But recent measurements \cite{cdf2} of the direct $J/\psi$ cross-section
(i.e. not coming frm $\chi$ decays) show that the theoretical
estimates are a factor 30-40 smaller. It has been suggested \cite{cgmp}
that a colour octet component in the $S$-wave production coming from
gluon fragmentation as originally proposed in Ref.~\cite{brfl}, can
explain this $J/\psi$ anomaly. The value of the colour-octet 
matrix-element is fixed by normalising to the data. 
The colour-octet contribution to $S$-state production
has also been invoked \cite{brfl} to explain the large $\psi^{\prime}$ 
cross-section measured by CDF \cite{cdf}, but there can be a large
contribution to this cross-section coming from the
decays of radially excited $P$-states \cite{psip}.

It is important to examine the implications of these two new
physics aspects of quarkonium production $viz.,$ fragmentation 
and the colour-octet contributions, for $J/\psi$ and $\psi^{\prime}$
production in other processes. Important tests of fragmentation 
at the Tevatron can be made by studying $J/\psi + \gamma$ production
at large $p_T$ \cite{psigam} and in $J/\psi$ pair production
at the Tevatron \cite{bfp}. Similarly, 
independent tests of the $S$-state colour octet enhancement are
important and it has been suggested \cite{sugg} that production
of quarkonia in $e^+e^-$ collisions or the measurement of the
polarisation of the $\psi^{\prime}$ \cite{bene1} can provide 
stringent tests of the colour-octet mechanisms.
We discuss, in the following, results from the analyses of
quarkonium production at HERA, LHC and in fixed-target experiments.

$J/\psi$ production at large-$p_T$ in $ep$ collisions at HERA has
been studied \cite{ours1}. Only inelastic events are considered: this is 
obtained by making a cut on the inelasticity parameter, $z$, defined as
\begin{equation}
z={p_{\psi} \cdot p_p \over p_{\gamma} \cdot p_p}.
      \label{e5}
\end{equation}
To ensure inelastic production of $J/\psi$, an
upper cut on $z$ is used, so that $z$ is sufficiently smaller
than 1.
The fusion contribution to the photoproduction of $J/\psi$ in the
colour-singlet model \cite{berjon} comes from photon-gluon
fusion. The next-to-leading order corrections to this process 
\cite{kramer} are in reasonable agreement with the data on
integrated inelastic cross-sections from HERA.
The integrated cross-sections are, however, insensitive to the
fragmentation contributions, and $p_T$ distributions need to be
studied to get a handle on the fragmentation contributions.

\begin{figure}
\vskip 9.5in\relax\noindent\hskip -0.8in\relax{\includegraphics{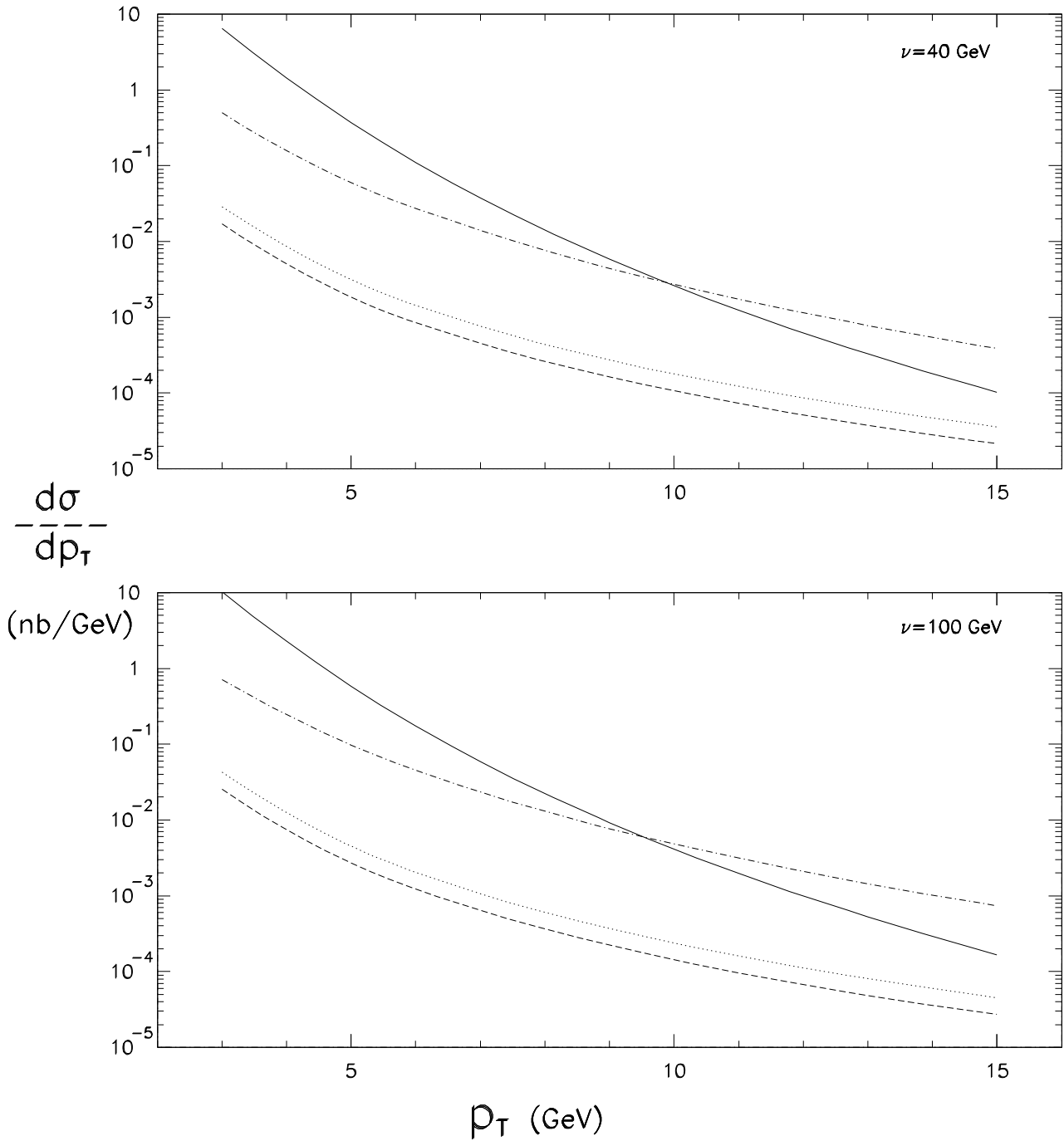}}

\vspace{-35ex}
\caption{$d\sigma/dp_T$ (in nb/GeV) for inclusive $J/\psi$ production
at HERA for photon energy $\nu=40$~GeV
(upper figure) and $\nu=100$~GeV (lower figure). The solid line
represents the fusion contribution
and the dashed-dotted line the charm quark fragmentation contribution.
The dotted and dashed lines 
represent the gluon fragmentation contributions with and without a 
colour-octet component for the $S$-state. 
The cut on the inelasticity parameter, $z$, is $0.1 < z <0.9$.}
\end{figure}

In addition to the above fusion process, we consider the contributions 
from the fragmentation of gluons and
charm quarks. The gluons are produced via the process
$\gamma + q \rightarrow q + g$,
whereas the charm quarks are produced via 
$\gamma + g \rightarrow c + \bar c$.
In principle, at HERA energies we can also expect
contributions from $B$-decays but these turn out to be dominant
at values of $z \le 0.1$ \cite{mns}, and can, therefore, be
eliminated by a lower $z$ cut.
We have computed the cross-sections for $\nu=40$ and 100~GeV, 
We use the cuts $0.1 \le
z \le 0.9$, as used in the ZEUS experiment at HERA \cite{zeus}. 
We find that the
fusion contribution, shown by the solid line in Fig.~1, is dominant
at low $p_T$, but the charm quark fragmentation contribution (shown
by the dashed-dotted line) becomes important for values of $p_T$ 
greater than about 10~GeV. The gluon fragmentation contribution 
(shown by the dashed line in the figure) is smaller by over an 
order of magnitude throughout the range of $p_T$ considered. 
The charm fragmentation subprocess is gluon-initiated 
and, hence, it dominates over the gluon fragmentation process. 
An experimental study of $p_T$ distributions at HERA will provide
us with the first direct measurement of the charm quark fragmentation
functions. To enhance the fragmentation contribution, it is efficient 
to use a stronger upper cut on $z$. 

At the LHC, we expect gluon
fragmentation to be the most important source of charmonium
production at large $p_T$, 
We have computed \cite{ours2} the cross-sections for the planned LHC energy
$\sqrt{s}=14$~TeV. In Fig.~2, we present the $J/\psi$ cross-section 
for a rapidity coverage $-2.5 \le y \le 2.5$. 

\begin{figure}
\vskip 8in\relax\noindent\hskip -0.8in\relax{\includegraphics{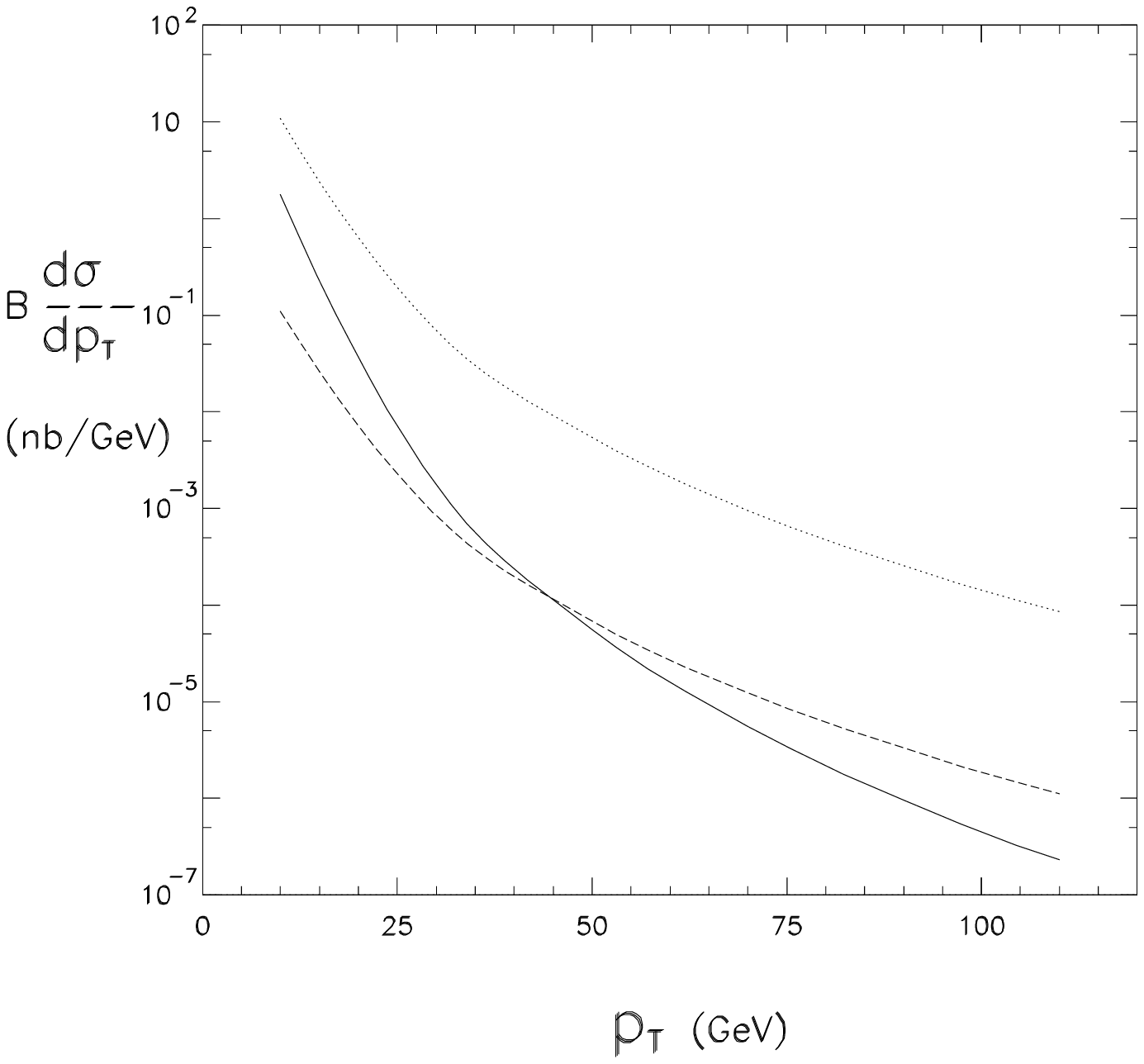}}

\vspace{-20ex}
\caption{$Bd\sigma/dp_T$ (in nb/GeV) for $J/\psi$ production at
14~TeV c.m. energy with $-2.5 \le y \le 2.5$. The
solid, dashed and dotted lines represent the fusion, charm quark
fragmentation and gluon fragmentation contributions.}
\end{figure}

We find that the cross-section for $J/\psi$ production is completely
dominated by the gluon fragmentation contribution (shown by the
dotted line in Fig.~2) and is larger than the fusion contribution
and the charm-quark fragmentation contribution by two orders of magnitude.
The cross-section for $J/\psi$ production is large and even at
$p_T=100$~GeV, the cross-section is as large as 0.1pb. 
For values of $p_T$ so much larger
than the charm quark mass, the fragmentation picture becomes 
exact and the experimental measurement of the $J/\psi$ cross-section 
will, therefore, be a crucial test of the fragmentation picture. 
The LHC measurement will also be a test of the magnitude
of the colour-octet contributions.

An important check of the colour-octet mechanism is provided by the 
comparison of the theory with $p_T$-integrated forward hadroproduction
cross-sections from fixed-target $p-N$ and $\pi - N$ experiments 
\cite{ours3}. A similar analysis of fixed-target hadroproduction
data has also been done in Ref.~\cite{bene2}.
The octet production cross-sections for the direct $J/\psi$
and for the $\chi$ states are needed for this comparison.

The octet cross section for $\jpsi$ is given by \cite{flem1}
\begin{equation}\begin{array}{rl}
     \sigma^8_{\jpsi}\;=\;
   \int_{\sqrt\tau}^1 {dx\over x} g_{\scriptscriptstyle P}(x)
                     g_{\scriptscriptstyle T}(\tau/x)
       {\displaystyle5\alphas^2\pi^3\over\displaystyle48m^5}&\mojpsb\\
       &+
   \lbrack \sum_f\int_{\sqrt\tau}^1 {dx\over x} q^f_{\scriptscriptstyle P}(x)
                     \bar q^f_{\scriptscriptstyle T}(\tau/x)
        +(P\leftrightarrow T) \rbrack
    {\displaystyle\alphas^2\pi^3\over\displaystyle54m^5}\mojpsa.
\end{array}\label{psi}\end{equation}
For the octet contributions for the $\chi_c$ states, the only 
cross-section required is
\begin{equation}
   \sigma^8_{\chi_J}\;=\;
   \lbrack \sum_f\int_{\sqrt\tau}^1 {dx\over x} q^f_{\scriptscriptstyle P}(x)
                     \bar q^f_{\scriptscriptstyle T}(\tau/x)
        +(P\leftrightarrow T) \rbrack
       {\displaystyle\alphas^2\pi^3\over\displaystyle54m^5}\mochij.
\label{chi}\end{equation}

The colour-octet matrix elements $\mochio$ and $\mojpsa$ have been 
extracted from the hadroproduction rates at the Tevatron \cite{cho2}. 
The remaining combination of matrix elements has been extracted from
photoproduction data \cite{flem2}. We use this value
\begin{equation}
   \mojpsb\;=\;0.020\pm0.001\ \ {\rm GeV}^3
\label{value}\end{equation}
in the calculations reported here\footnote{The individual matrix
elements appearing in this linear combination are, however,
substantially smaller than those determined by a comparison to 
the Tevatron large-$p_T$ data.\cite{cho2}}.

\begin{figure}
\vskip14truecm
\includegraphics{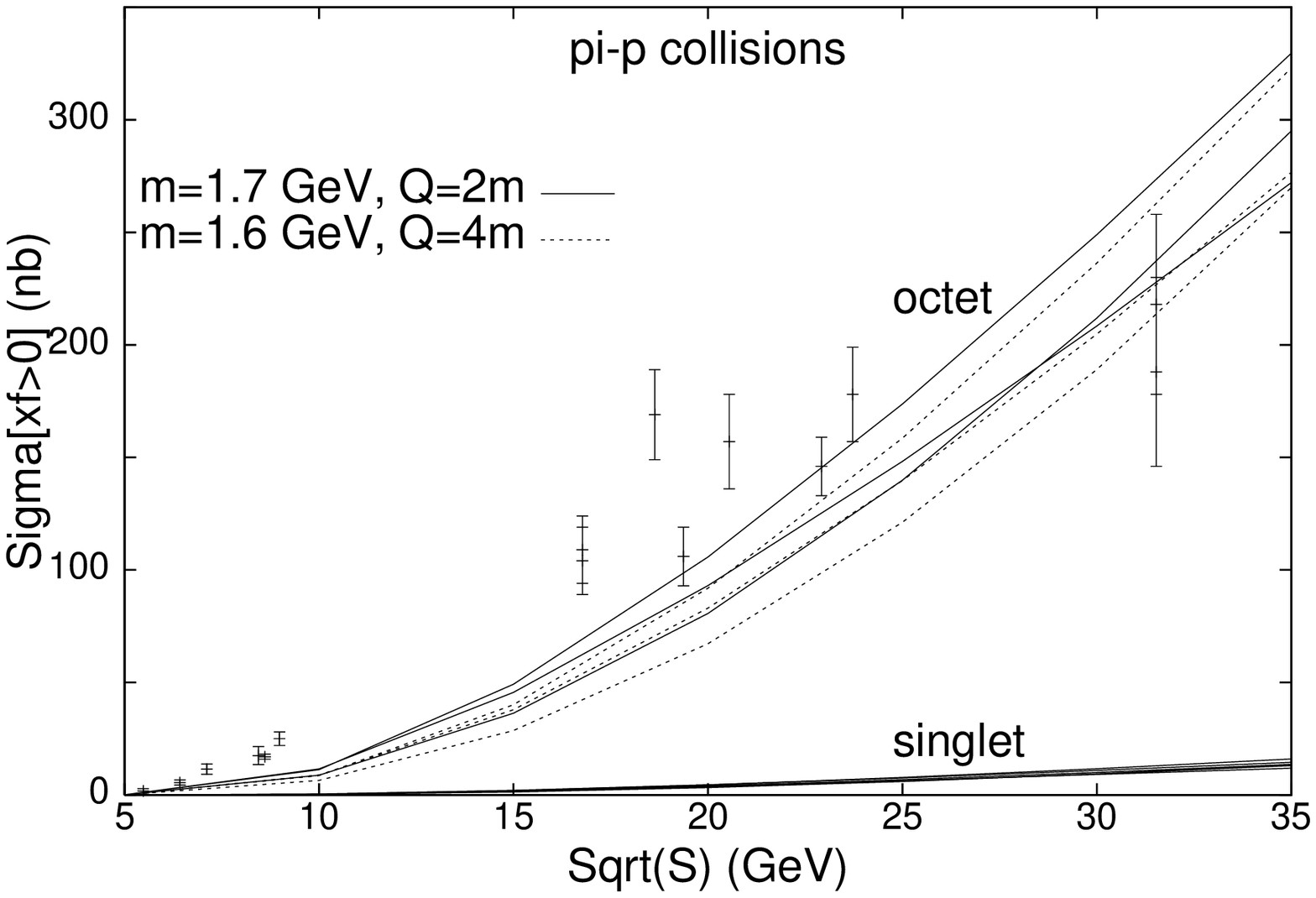}
\includegraphics{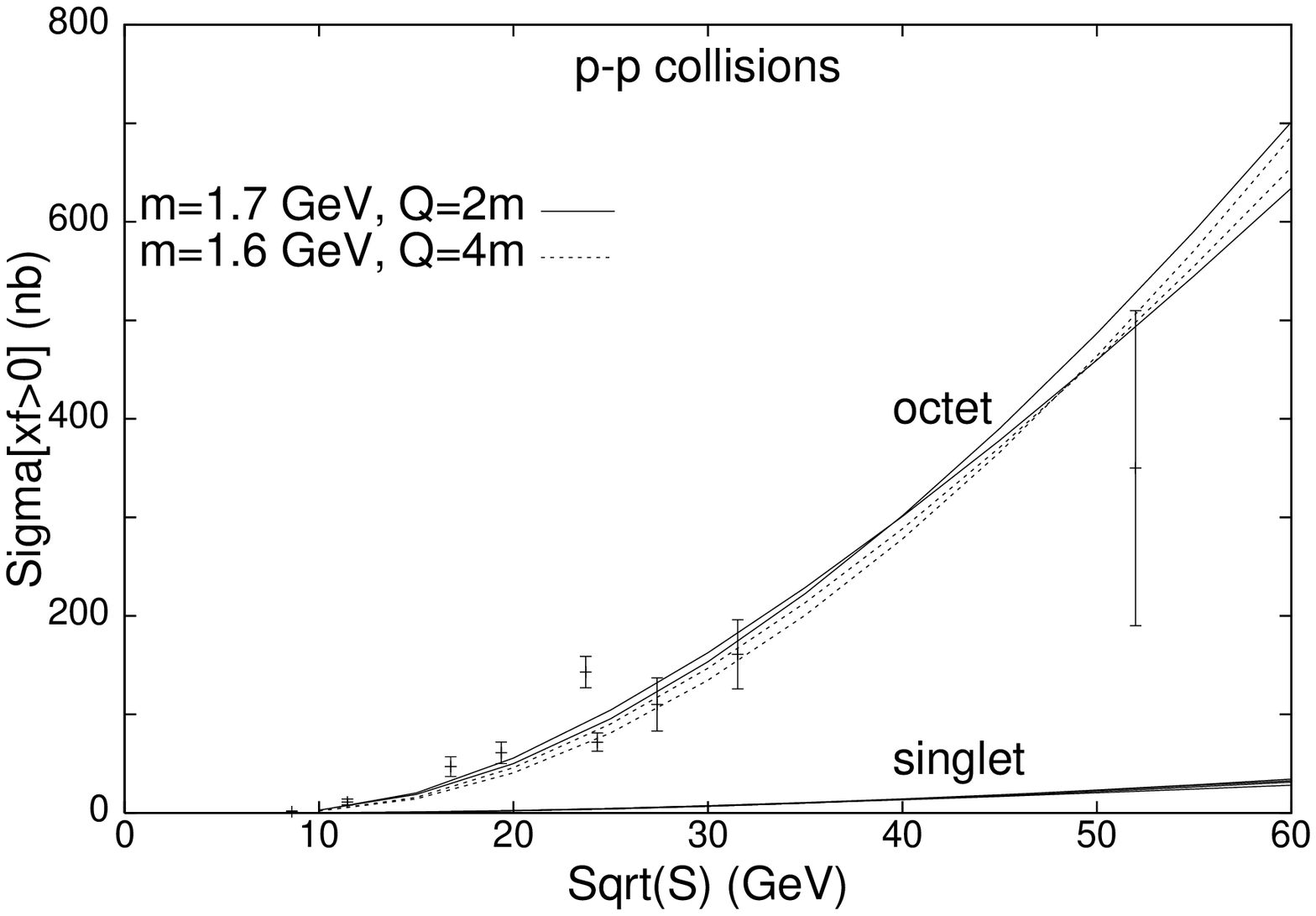}
\caption[dummy]{The colour-octet model predictions for integrated forward
  $\jpsi$ hadroproduction cross sections as a function of the CM energy.
  For $pp$ collisions, the two curves for each choice of $m$ and scale $Q$
  are for the structure functions MRS ${\rm D}-'$ and GRV LO. For $\pi p$
  collisions the three sets of structure functions are
  MRS ${\rm D}-'$ for proton and SMRS 1 for $\pi$, MRS ${\rm D}-'$
  for proton and SMRS 3 for $\pi$, GRV LO for both. Note that the
  colour-singlet model predictions lie far below the data.}
\label{fig}\end{figure}
Our results are shown for the two choices
$m=1.7$ GeV and a scale of $2m$, as well as $m=1.6$ GeV and a scale of $4m$.
With these inputs, we find that the $\sqrt S$ dependence of
the integrated forward $\jpsi$ production rates, for both $pp$ and $\pi p$
collisions, are described rather well by the model (see Figure \ref{fig}).
We would like to emphasise that there are no free parameters in this
calculation.

In summary, we have discussed two new aspects of quarkonium physics:
fragmentation and the colour-octet mechanism. We have further discussed
how inelastic large-$p_T$ photoproduction of $J/\psi$ at HERA and 
large-$p_T$ hadroproduction of $J/\psi$ production at LHC will 
provide information on charm fragmentation and gluon fragmentation 
mechanisms, respectively. The energy dependence of $p_T$-integrated
forward cross-sections from fixed-target $p-N$ and $\pi - N$ experiments
are very well reproduced after the colour-octet Fock components have
been taken into account. This is a clear indication of the neccessity
of including the octet components for a complete description of quarkonium
processes.

\end{document}